# RADIO SETI OBSERVATIONS OF THE INTERSTELLAR OBJECT 'OUMUAMUA

Gerry R. Harp, Jon Richards, Peter Jenniskens, Seth Shostak, Jill C. Tarter

Center for SETI Research, SETI Institute, 189 Bernardo Ave., Mountain View, CA 94043

**Abstract**
Motivated by the hypothesis that 'Oumuamua could conceivably be an interstellar probe, we used the Allen Telescope Array to search for radio transmissions that would indicate a non-natural origin for this object. Observations were made at radio frequencies between 1-10 GHz using the Array's correlator receiver with a channel bandwidth of 100 kHz. In frequency regions not corrupted by man-made interference, we find no signal flux with frequency-dependent lower limits of 0.01 Jy at 1 GHz and 0.1 Jy at 7 GHz. For a putative isotropic transmitter on the object, these limits correspond to transmitter powers of ~~30 mW~~ 10 W and ~~300 mW~~ 100 W, respectively. In frequency ranges that are heavily utilized for satellite communications, our sensitivity to weak signals is badly impinged, but we can still place an upper limit of ~~10 W~~ 3 kW for a transmitter on the asteroid. For comparison and validation should a transmitter be discovered, contemporaneous measurements were made on the solar system asteroids 2017 UZ and 2017 WC with comparable sensitivities. Because they are closer to Earth, we place upper limits on transmitter power to be 0.1 and 0.001 times the limits for 'Oumuamua, respectively.

*Keywords: asteroids: individual ('Oumuamua), comets, interstellar, SETI*

## Introduction

The discovery of object 1I/2017 U1 ('Oumuamua) by the PanSTARRS1 telescope system in October 2017 [1] has generated interest both professional and popular. Its hyperbolic trajectory indicates that it is not bound to our solar system, and has presumably been ejected from the environs of another star. At the very least, objects like 'Oumuamua provide an opportunity to directly learn something about the physical conditions in other stellar environments.

'Oumuamua passed within 0.25 AU of the Sun, and if – as initially thought – this is a comet, then outgassing should have occurred. No comet coma was observed, and consequently the object was reclassified as an asteroid. More recently it was found from careful analysis of astrometric data that 'Oumuamua experienced non-gravitational forces, suggesting some amount of outgassing, albeit much less than that of typical Oort cloud comets [2]. It has been suggested that 'Oumuamua could still be an icy object, but that hardening of an outer layer might have been caused by long-term exposure to cosmic rays [3]. Oort cloud comets, however, were also exposed to cosmic rays for 4.55 Ga and did not develop such a hard layer.

The object has a reddish tint typical of primitive asteroids and comet nuclei [1]. Brightness variations indicate a tumbling motion with a period of 7 – 8 hours (e.g., [4]). 'Oumuamua, with an estimated size of several hundred meters, has a large axial ratio – perhaps as large as 10:1 [1]. Such a cigar-shaped morphology has never been found for a solar system asteroid. Several suggestions have been made as to how a natural object can be this elongated (e.g., [5]).

While all measures made of 'Oumuamua are consistent with it being either an asteroid or a comet, another scenario has attracted attention. The object's unusual shape, coupled with its passage through the near-center of our solar system and passing only 0.096 AU from Earth's orbit (could it have been targeted?), raises the possibility that 'Oumuamua is an interstellar probe, constructed either by fabrication or by hollowing out an existing object. This provocative idea gains much of it's following because of its similarity to the story premise of the popular novel by Arthur C. Clarke, "Rendezvous with Rama" [6].

Consequently, we undertook observations of this object with the Allen Telescope Array beginning November 23, 2017. We also observed two other normal Sun-bound asteroids, 2017 UZ and 2017 WC, to better characterize the background radiation. If evidence for a transmitter were found in 'Oumuamua observations, direct comparisons would be made for the same frequency on the other asteroids as a means of confirmation.

**Methods**

The Allen Telescope Array (ATA) was used over the course of eight days from November 23 – December 5, 2017 to search for non-natural narrow-band radio emissions. 'Oumuamua at the time was between 1.43 and 1.89 AU from the Sun (Fig. 1). The array consists of 42 antennas, each 6.1 m in diameter, and has been described elsewhere [7]. Nineteen antennas were used for the present experiment, specifically antennas 1, 2, 4, 5, 6, 10, 11, 13, 14, 16, 19, 24, 25, 29, 30, 36, 38, 40, and 42.

We searched for moderately wide bandwidth signals >100 kHz. To this end, frequencies between 1 and 10 GHz were observed. We employed an imaging correlator, set up with an instantaneous bandwidth of 80 MHz and 100 kHz minimum resolution. Following a procedure similar to that used in [8], frequency power spectra were generated by fitting a point source and background level (*MIRIAD* task *uvfit*) at the observation reference center, when placed in the direction of 'Oumuamua or another target. Observations lasted 10 minutes and were repeated at least twice at each frequency on different days. The absolute flux scale was set by contemporaneous observations of quasars, either 3c273 or 3c123, with frequency dependent assumed fluxes at each observation frequency interpolated from published Very Large Array calibration data [9]. Special care and improved algorithms resulted in sensitivity levels here that are much lower than in an earlier paper [8]. If transmissions were found that possibly originated on 'Oumuamua, two-dimensional images were generated using the correlator data as a spot check of flux localization. For a comparison of images of true emission versus interference signals, see [10].

**Results**

Correlator observations would reveal any wideband (> 100 kHz) signals coming from the direction of the object. A plot of the raw fits to the observations is given in Fig. 2, which summarizes 90,000 fitted points. This implies a great deal of over-plotting in this small figure. We include all valid observations so that all outlier points are visible. This sets a strong upper limit to the fitted flux.

The most striking features in Fig. 2 are tall bunches of points occurring over specific frequency ranges. In all cases, the frequency range of these bunches corresponds to heavily occupied regions of the frequency spectrum, mostly in the ranges of satellite communication downlinks. Thus, man-made interference dominates the uncertainty in our measurements of flux in those regions. Due to background noise, fitted fluxes are sometimes negatively valued, and indeed the graph is highly symmetrical about the 0 Jy line. The intrinsic telescope sensitivity was typically much better than the measurement uncertainty caused by man-made interference.

More meaningful than the raw observed flux would be to estimate the transmitter power at each asteroid that could have been detected versus transmission frequency. Knowing the range to each asteroid, we can plot upper limits of transmitter power associated with observations on each asteroid (Fig. 3). 2017 WC and 2017 UZ were at about $2.6 \times 10^6$ and $7.5 \times 10^7$ kilometers, respectively, while the average range of 'Oumuamua was $2.5 \times 10^8$ kilometers. Because the sensitivity is sometimes limited by interference, we estimate the uncertainty in flux measurements using the variance of the data itself (σ estimated from 100 points nearest the observation frequency). We estimate the 7.5 σ RMS power variation of the fitted data[1] for each asteroid (blue lines in Fig. 3). By comparison of the graphs in Figure 3, we find these curves envelope all measured data points, and serve as estimates of the minimum transmitter flux that would have been detectable.

Depending on the interference environment, the upper limit placed on a bandwidth-matched transmitter varies by more than 3 orders of magnitude from a fraction of a Watt 10 Watts to approximately 10 Watts 3000 Watts. For comparison, we note that a typical cell phone transmitter produces a peak radiated power of about 1 Watt. It should be noted that throughout this paper and above, the power limits are quoted for continuous signals. Transmitter power plots were also generated for the two solar system asteroids at the middle and bottom of Fig. 3. Since these asteroids are substantially closer to Earth, we can set more stringent bounds on isotropic transmitter power for those cases. For the closest asteroid 2017 WC, our observations rule out any transmitters (bandwidth > 100 kHz) with powers greater than ~~10 mW~~ 0.3 W.

---

[1] It is important to notice that $\sigma_{fit}$ = RMS deviation of fitted power quoted here is smaller than and essentially unrelated to the average noise power ($\sigma_{Radiometer}$ = RMS voltage * Jy/Volts) computed from the radiometer equation. To understand this difference, notice that our fitting routine fits both a point and a constant background value to approximately 1000 measures (beams) in the observed image data, hence $\sigma_{fit}$ represents the peak variation *after* subtraction of a constant background. This explains why the data in Fig. 2 are symmetric about zero power instead of symmetric about the average noise power level. Since $\sigma_{fit}$ is estimated *directly* from the data in Fig. 3, this value of sigma is consistent with the data and for the plots of Fig. 3. By comparison, we estimate the value of the more conventional radiometer standard deviation to be $0.1 < \sigma_{Radiometer} < 1$ Jy, for the ATA in this application.

It is expected that the apparent strength of interfering sources will vary depending on pointing, so the flux curves on 2017 UZ and 2017 WC do not track with 'Oumuamua or one another exactly. This can be observed in Fig. 3, which translates the observed frequency intensities to the isotropic power of a radio transmitter on the asteroid that would be required to give the observed signal. While the individual peaks vary in strength from differences in pointing, the observed frequencies are the same, as expected for interference.

Some images were produced as spot checks of flux localization at a variety of frequencies, mostly using the 'Oumuamua data. In all cases, there was no indication of a point source at image center.

**Discussion**

Previous work has tested the hypothesis that 'Oumuamua is emitting radiation at radio frequencies. A short note [11] describes observations made with the Green Bank Telescope over much of the same frequency range as reported here (Dec. 13-27, 2017). For a 3 Hz bandwidth transmitter, the upper limit on isotropic radiated power was ~80 mW. At lower frequencies (72 - 102 MHz), observations with 10 kHz bandwidth placed an upper limit of 840 mW for an isotropic transmitter [12] (Nov. 8, 2017).

Our new observations were made during a different epoch (Nov. 23, 2017 – Jan. 1, 2018, with the possibility that transmissions were intermittent), search over a bandwidth of 100 kHz and higher, and includes comparisons with two in-solar-system asteroids. Our lower limits to transmitter powers of ~~30 mW~~ 10 W and ~~300 mW~~ 100 W at 1 and 7 GHz, respectively, are complimentary to previous observations.

These limits refer to an isotropic transmission, which may be less likely than either beaming a signal towards Earth, or towards some star system that might have been the origin point of any probe. In the former case, the limits on the transmitter power would be lower by a factor equal to the gain of the antenna.

These limits are many orders of magnitude lower than those typically quoted in SETI observations of stellar systems, and constitute a sensitive test for signals that would betray 'Oumuamua as a deliberate probe.

The present survey is a prototype of a radio SETI search that focuses on asteroids bound to our solar system. It has long been recognized [13] that some asteroids confined to our solar system could be space probes manufactured elsewhere. Such probes might be put into a stable orbit as a means of monitoring our solar system for evidence of technology, perhaps communicating results back to its originators. Our observations demonstrate how very weak transmitters attached to asteroid-like objects might be detected.

**Conclusions**

We made a careful search for artificial emissions from three nearby asteroids, including the interstellar object 'Oumuamua. No evidence for such emissions was found in the observational data. We place upper bounds on the power output of an isotropic transmitter located at each asteroid. This upper bound is highly frequency dependent, due to unfortunate interference by man-made signals, and also dependent on the distance to the asteroid. Frequency dependent upper bounds for transmitter power are estimated for 'Oumuamua ~~(0.1 – 10 W)~~ (10 - 3000 W), for 2017 WC ~~($10^{-5}$ – $10^{-2}$ W)~~ (3 mW – 1 W) and 2017 UZ ~~($10^{-2}$ – 1 W)~~ (10 – 300 W). Besides its value as a study of the extremely interesting interstellar object, 'Oumuamua, we also prototype the analysis pipeline for a future search for space probes bound to our solar system.

**Acknowledgements** - We thank those that have supported the Allen Telescope Array over the years. PJ is supported by NASA's SSO program.

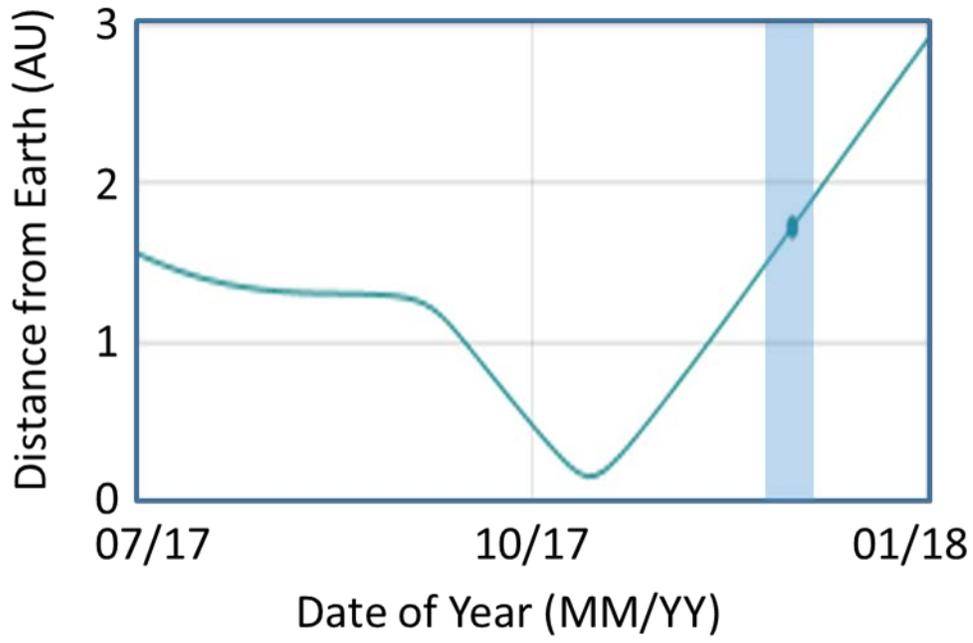

Figure 1: A plot of the range of 'Oumuamua from Earth over the specified dates. The rectangle indicates the time period when ATA observations were made.

Figure 2: Plot of point source fits to the 'Oumuamua position at 100 kHz intervals from 1-10 GHz.

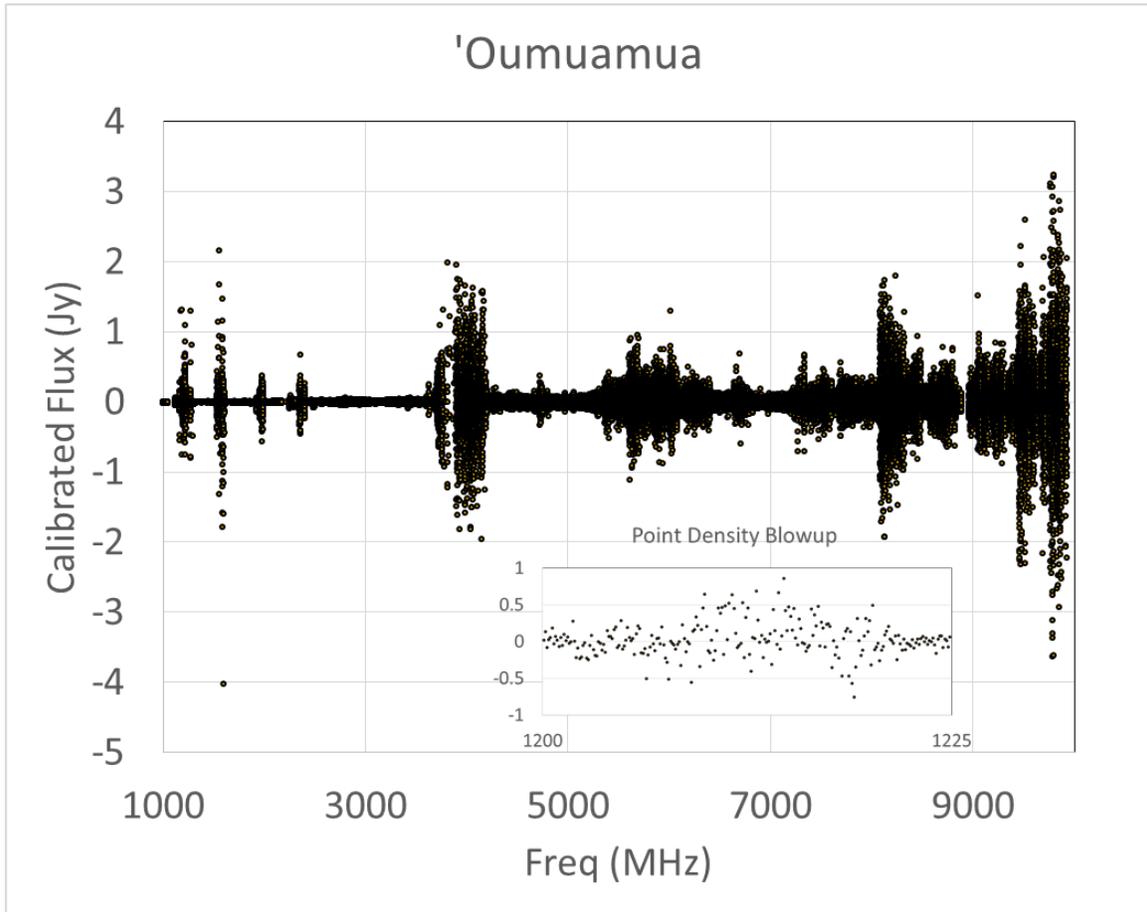

Figure 3: **REVISED FIGURES:** Power of an isotropic transmitter located at each asteroid to generate the observed flux. The line corresponds to the 7.5 σ upper limit. The values in these figures are $\sqrt{100,000}$ times larger than in the erroneous, previously published version.

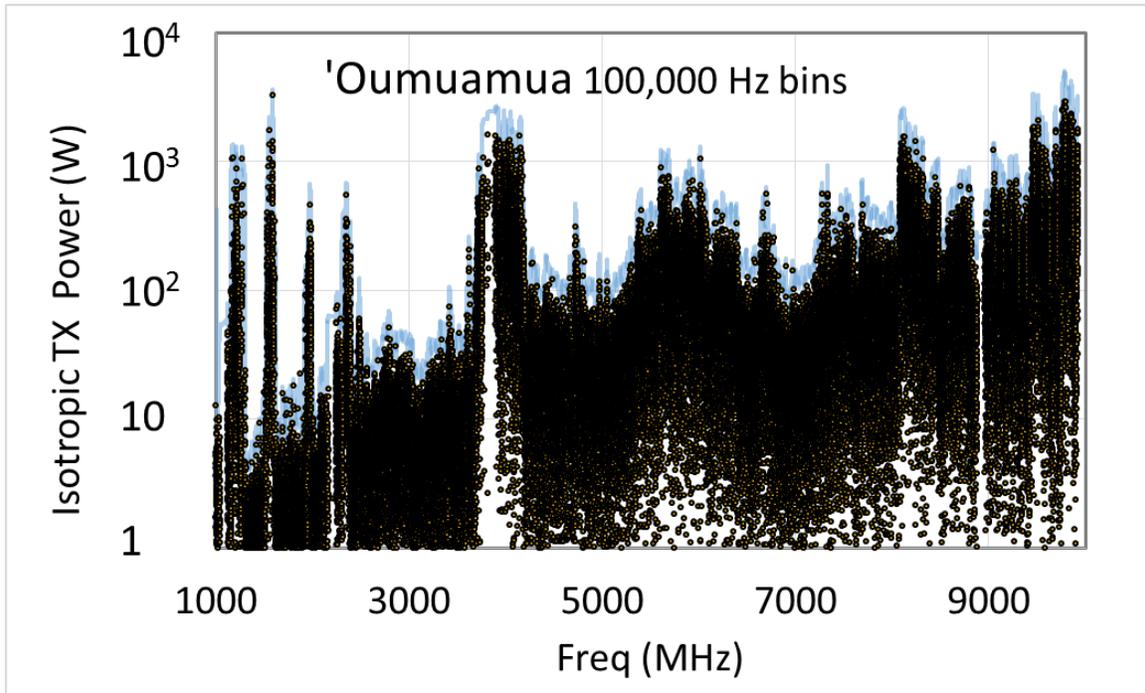

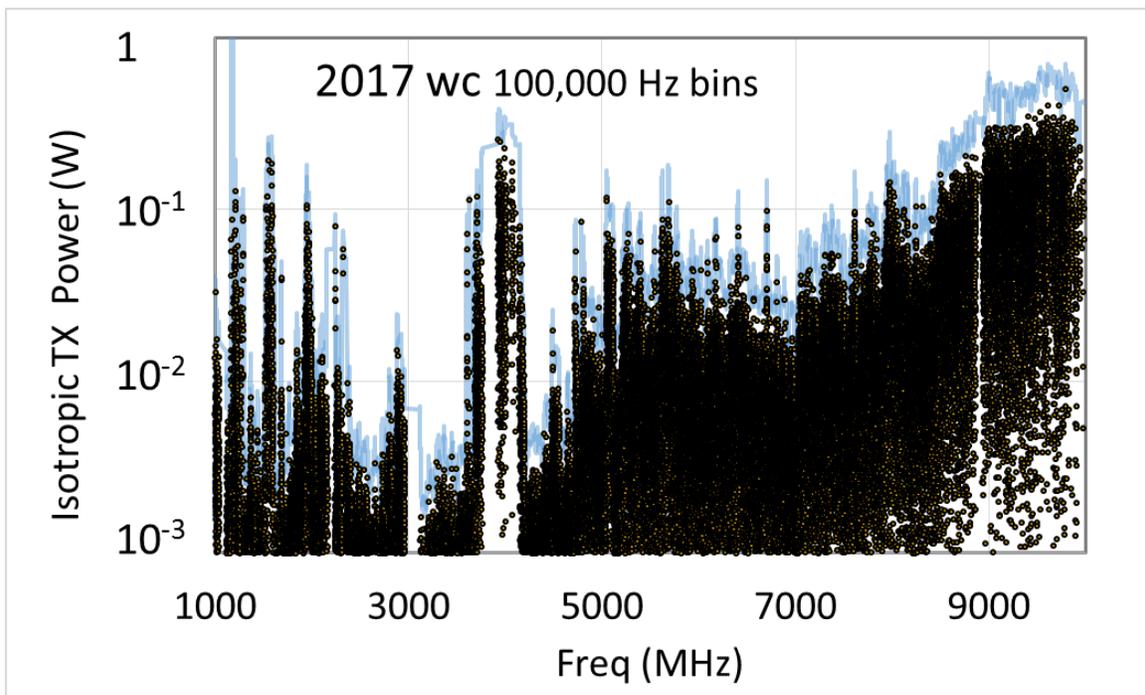

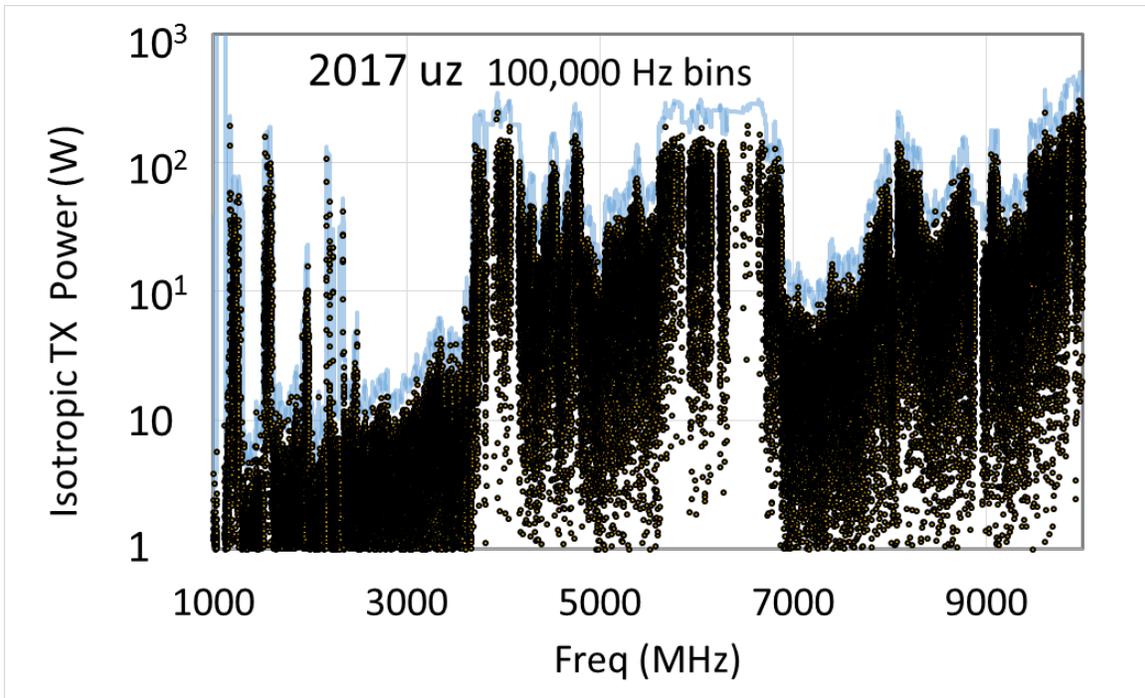

**Erratum:** Correction to calculation of transmitter power on the asteroids that corresponds to the non-observation of flux on these sources at a level of 7.5 $\sigma$.

An unfortunate mistake crept into the calculation of transmitter powers by a factor of the square root of the bandwidth or $\sqrt{100,000} = 316$. Although the flux measures of Figure 2 are unaffected, the transmitter powers in Figure 3 should all be revised upward by this amount. This version of the document displays the revised transmitter scales. The numbers in the abstract and text have been changed to reflect this correction, in red.

G. R. Harp
11-Jan-2019